\documentclass[12pt]{article}

\usepackage{arxiv}

\usepackage[utf8]{inputenc} 
\usepackage[T1]{fontenc}  

\usepackage{booktabs}   
\usepackage{amsfonts}   
\usepackage{nicefrac}      
\usepackage{microtype}     
\usepackage{lipsum}
\usepackage{fancyhdr}      
\usepackage{graphicx}      
\graphicspath{{media/}}   
\usepackage{amsmath,latexsym,amssymb}
\usepackage{amsthm}
\inputencoding{utf8}

\input{Qcircuit}

\newtheorem{Theorem}{Theorem}
\newtheorem{Lemma}{Lemma}
\newtheorem{Property}{Property}
\newtheorem{Definition}{Definition}

\newcommand{\Endproof}{\hfill$\Box$ \\}

\title{Hybrid classical-quantum text search based on hashing}
\author{Farid Ablayev \And Marat Ablayev \And Nailya Salikhova}

\begin{document}

\maketitle

\renewcommand{\refname}{\small {References}} \renewcommand{\abstractname}{Abstract} 
\renewcommand{\figurename}{Pic.}
\renewcommand{\proofname}{ {\hskip\parindent \bf Proof. }}

\begin{abstract}

The paper considers the problem of finding a given substring in a text. It is known that the complexity of a classical search query in an unordered database is linear in the length of the text and a given substring. At the same time, Grover's quantum search provides a quadratic speedup in the complexity of the query and gives the correct result with a high probability.

We propose a hybrid classical-quantum algorithm  (hybrid random-quantum algorithm to be more precise),  that implements Grover's search to find a given substring in a text. As expected, the algorithm works a) with a high probability of obtaining the correct result and  b) with a quadratic query acceleration compared to the classical one.

What's new is that our  algorithm uses the uniform hash family functions technique. As a result, our algorithm is much more memory efficient (in terms of the number of qubits used) compared to previously known quantum algorithms.

\end{abstract}

\section{Introduction.}

 Quantum search algorithms and its generalizations \cite{grover1996fast,gilliam2020optimizing, brassard2002quantum,reitzner2012quantum, wolf2019}  are part of a group of quantum algorithms that are of great interest in computer science.

The using of quantum computers can significantly reduce the computation time for models with a large number of weights, or those requiring an exponentially growing number of combinations. We presented a review of some results in the field of quantum approaches to classification and information retrieval in the papers \cite{ablayev2019quantum, ablayev2019quantum2}.

The task of finding occurrences of a given substring in a text is important problem  of information retrieval. It occurs in a wide range of applications, namely, in text editors, in search robots, spam filters, bioinformatics, etc. A large number of algorithms (deterministic and probabilistic) for solving the search problem have been developed over the past few decades. Quantum algorithms for solving the search problem have been developed in the last two decades.

\paragraph*{ Problem.}
Given a binary $N$ length  sequence 
\[string=b_1\dots b_N \] 
and  a binary $m$ length  sequence $w$,  $m<N$. It is required to find the index of the occurrence of the substring $w$ in the text $string$. Namely, it is required to find an index $k$ such that $w=b_{k}\dots b_{k+m-1}$.

\paragraph*{Related work.}
The known Knuth-Morris-Pratt's \cite{knuth1977fast} classical algorithm (1977) solves the problem in linear time $O(m + n)$.

In the early 2000s, a quantum algorithm \cite{ramesh2003string} for searching for a given substring in a text was presented. It allows to get quadratic search speed-up.
It returns one of the indexes of the occurrence of the searched substring. The probability of getting the correct answer is strictly greater than 1/2. The query complexity (this measure of complexity is sometimes associated with time complexity) of an algorithm is $O(\sqrt{n}\log{\sqrt{\frac{n}{m}}}\log{m} + \sqrt{m}\log^2{m})$. The authors \cite{ramesh2003string} do not estimate the memory complexity (number of used qubits) of their algorithm. 
An analysis shows that the \cite{ramesh2003string} algorithm requires $O(\log{n} + m)$ qubits.

\paragraph*{ Our contribution. } We present a quantum algorithm for finding a pattern in a string. The algorithm is based on Grover's search algorithm and hashing technique. It is known that the hashing method can significantly save space and, as a rule, the time required for searching in databases. The results of the paper  demonstrate the potential of a universal hashing method for building quantum search algorithms. 

To simplify the presentation, not to clutter the idea with technical details, consider the case when it is known in advance that the desired substring occurs in the text exactly once. 

More precisely, we propose a hybrid random-quantum search algorithm $\cal A$ that searches a text for a substring and has the following characteristics:
\begin{itemize}

\item The $\cal A$ algorithm produces a result with a high probability of obtaining the correct answer.

\item The $\cal A$ algorithm is based on Grover's search. This search is presented in the paper as an auxiliary algorithm ${\cal A}1$ and requires $O(\sqrt{n})$ query steps.
\item The $\cal A$ algorithm exponentially saves (compared to the \cite{ramesh2003string} algorithm) the number of qubits relative to the parameter $m$ -- the length of the substring. Namely, the algorithm requires $O(\log { n}+\log{ m })$ qubits for his work.
\end{itemize}

The main idea of the paper is the use of hashing technique  to save space complexity  in quantum search. The ${\cal A}$ algorithm is based on a certain universal family of hash functions. The ${\cal A}2$ algorithm is a generalization of the result to a general universal hash family of functions.

The result of the work is organized as follows. In the next section (Section 2), we present the basic model of the quantum search algorithm, basic notation and definitions. In Section 3 we present the main result. We first present the auxiliary quantum procedure ${\cal A}1$. Next, we present a quantum search algorithm ${\cal A}$ based on the hashing technique. Theorem \ref{th_a2} is an analysis of the ${\cal A}$ algorithm. In Section 5, we present the ${\cal A}2$ algorithm, a generalization of the ${\cal A}$ algorithm. 

The preliminary version of the article was published in Russian  \cite{UchZap_2020}.

\section{Preliminaries for quantum query algorithm }

We refer the reader to the  \cite{wolf2019,ambainis2017} for an introduction to the basics of quantum query algorithms and the state of research in this area. Here we  define the query model of computations in a way that is convenient for us. We use the notation defined in detail in \cite{ablayev2019quantum, ablayev2019quantum2}. 

The operations applied to quantum $s$-qubit states $\ket{\psi}\in {({\cal{H}}^2)}^{\otimes s }$ are mathematically expressed using unitary operators:
\begin{equation}\label{operator}
\ket{\psi'} = U\ket{\psi},
\end{equation}
where $U$ is a $2^s\times 2^s$ unitary matrix.

\paragraph{Query model.}   For a finite sets $X$ and $Y$ let 
\[ g: X\to Y\]
be a discrete function. A quantum query algorithm $\cal A$ computing the function $g$ begins in a quantum state 
$|\psi_{start}\rangle$ and  applies  a sequence of operators
\[  {\cal O}_X, {\cal U}, \dots , {\cal O}_X, {\cal U},\]
The operator ${\cal O}_X$ depends on the input $X$. The quantum community calls the ${\cal O}_X$ operator an ``oracle''. Application of the oracle is called the query of the algorithm $\cal A$ to the initial data $X$. The operator $\cal{U}$ does not depend on $X$.

The algorithm computes the value $g(\sigma)$ of the discrete function $g$ for $\sigma\in X$ if the initial state $|\psi_{start}\rangle$ goes to the final state
\[
\ket{\psi(g(\sigma))}=
{\cal U} {\cal O}_\sigma\cdots {\cal U} {\cal O}_\sigma  \ket{\psi_{start}}, 
\]
in the process of computing on the input value $\sigma$.
The final state allows extracting the value of $g(\sigma)$ as a result of measuring the state of $\ket{\psi(g(\sigma))}$.

\paragraph*{Search in an unordered database.} Quantum search algorithm in an unordered database of $n$ elements, in which there is exactly one element of interest to us (Grover's algorithm) is a special case of a group of query algorithms.
The following algorithm scheme underlies many generalizations.
\begin{enumerate}
    \item  Initialize a quantum system of $O(\log n)$ qubits into a $\ket{\psi_{start}}$ state containing information about the database. The $\ket{\psi_{start}}$ state is constructed so that each of the $2^{O(\log n)}$ basis quantum states represents the required information about $n$ database elements.
\item  The following $O(\sqrt{n})$ ``macro'' steps are performed on the $\ket{\psi_{start}}$ state
\begin{itemize}
\item The oracle ${\cal O}_X$ is used. It recognizes the basis state of interest to us and multiplies its amplitude by -1.
\item  The operator $\cal U$ performs the inversion by the average value over all amplitudes.
\end{itemize}
\end{enumerate}

``Macro'' Step 2 describes the key operation of quantum search. Each such step increases the amplitude of the basis state that represents the desired information. The number $O(\sqrt{n})$ of such steps maximizes the amplitude, and  the probability of extracting the required information from the resulting state $\ket{\psi_{final}}$ becomes close to 1.

\paragraph*{ Basic operations of qubits for quantum search.} 
The potential advantages of quantum algorithms, on which the results of this work are based, lie in the possibility of implementing quantum operators of dimension $2^s\times 2^s$ by basic operations based on a small number of order $O(s)$ qubits and in a small number of quantum computing steps (not a complex scheme implemented by basic elements).

Main operations are $I,X,Z,H$.

\begin{itemize}
    \item 
$I$ - identity operator.
\[ I=\left(\begin{array}{cc}1 & 0\\0 & 1\end{array}\right).\]

\item $X$ - is a NOT operator. It changes the state of the qubit from $\ket{0}$ to $\ket{1}$, and vice versa.

\begin{equation}\begin{array}{l}
X=\left(\begin{array}{cc}0 & 1\\1 & 0\end{array}\right).
\end{array} 
\end{equation}

\item $Z$ - amplitude sign reversal operator.
\begin{equation}\begin{array}{l}
Z =\left(\begin{array}{cc}1 & 0\\0 & -1\end{array}\right).
\end{array}
\end{equation}

\item $H$ - Hadamard operator. 
\begin{equation}\begin{array}{l}
H=\frac{1}{\sqrt{2}}\left(\begin{array}{cc}1 & 1\\1 & -1\end{array}\right).
\end{array}
\end{equation}

\end{itemize}

\paragraph*{ Characteristics of the ${\cal A}$ algorithm.} The characteristics of a quantum query algorithm are the used memory, the number of queries to the analyzed data, and  the probability of an error.

\begin{quote}
  
\item {\em Size complexity (Memory complexity).} The number $S({\cal A})$ of used qubits is a measure of the memory complexity of the quantum algorithm ${\cal A}$.  

 We denote by $S^{\cal A}(string,w)$ the number of qubits used by the $\cal A$ algorithm to solve the problem of finding the substring $w$ in the $string$.

Through $S^{\cal A}(n,m)$ we denote the maximum among the numbers $S^{\cal A}(string, w)$ over all $string$ and $w$ with parameters $N=|string|, m=|w|, n=N-m+1$.

\item {\em Query complexity.} The number $Q({\cal A})$ of queries (number of the oracle ap\-pli\-ca\-tions) is a ``query'' measure of the complexity of the quantum algorithm ${\cal A}$. Note that in \cite{ambainis2017}, one request to the oracle is testing one variable (one bit). Accordingly, in our work, one request to the oracle is implemented in the process of testing the entire string $w$ (hash values  of $w$). 

We denote by $Q^{\cal A}(string,w)$ the number of applications by the algorithm $\cal A$ of the  oracle operator. 

Through $Q^{\cal A}(n,m)$ we denote the maximum among the numbers $Q^{\cal A}(string, w)$ over all $string$ and $w$ with parameters $N=|string|, m=|w|, n=N-m+1$.

\item {\em Error probability.} 

We denote by $Er^{\cal A}(string, w)$  the probability of the following event. The algorithm $\cal A$ as a result of solving the problem of finding the substring $w$ in the text $string$ gives the number $k$ of the position in the text $ string$ such that $w_k\not = w$. 

Through
$Er^{\cal A}(n,m)$ we denote the maximum among the numbers $Er^{\cal A}(string, w)$ over all $string$ and $w$ with parameters $N=|string|, m=|w|, n=N-m+1$.

\end{quote}

\section{Algorithms for finding the index of occurrence of a substring in the text.}

In this section, we present a hybrid classical-quantum algorithm (more precisely, a hybrid  random-quantum algorithm) ${\cal A}$ for finding a substring $w$ of length $m$ in a binary text 

\[string=b_1\dots b_N. \]

We consider the simplest version of the problem: we assume that the required substring $w$ is guaranteed to occur exactly once in the text $string$.

The problem is  reduced to a problem of finding  word $w$ in a vocabulary as follows. Let  $n=N+1-m$. To simplify presentaion, we  assume below that the number $n$ (the number of substrings in the  $string$) is a power of 2. Denote by $V(string, m)$ 

a sequence  composed of all substrings of length $m$ of the  $string$
\[V(string,m) = \{w_0,\dots , w_{n-1} \},\]
where $w_k=b_{k+1}\dots b_{k+m}$ for $0\le k \le n-1$. 
We will call $V(string, m)$ a vocabulary.

Now the problem of finding the word $w$ in $string$ is represented as the problem of finding an index $k$  such that $w=w_k$ for the vocabulary $V(string, m)$.

\subsection{Auxiliary quantum procedure ${\cal A}1$.}

The quantum part of the algorithm is the following auxiliary quantum procedure ${\cal A}1$. Procedure ${\cal A}1$ starts with an initial quantum state $\ket{string}$ (quantum vocabulary) constructed from the vocabulary $V$ using the preprocedure $\cal P$.

{\bf Pre-procedure $\cal P$} generates  the initial state  (quantum vocabulary) $\ket{string}$ from vocabulary $V=\{v_0, \dots, v_{n-1}\}$ composed of binary words of length $l\ge 1$.  The $\ket{string}$ state has the following structure
\[
    \ket{string} =
    \frac{1}{\sqrt{n}}\sum\limits_{k=0}^{n-1}\ket{k}
    \otimes \ket{v_k}
    \otimes \ket{1}.
    \]
 
\paragraph{Description of the procedure ${\cal A}1$}
\begin{quote}
    \item {\bf Input:} 
Quantum state $\ket{string}$. Binary word $v$. 

\item {\bf Output:} The index $k$, which is interpreted as an index such that $v=v_ k$. 
\end{quote}

That is, ${\cal{A}}1(\ket{string},v)$ implements the mapping
\[
{\cal{A}}1 : \ket{string}, v \longmapsto k.
\] 

The following two macro steps, described below in operator form, are applied to the state $\ket{string}$ $\frac{\pi}{4}\sqrt{n}$ times. In the literature, these two macro steps are often referred to as the ``Grover iteration'' \cite{grover1996fast}. 

\begin{itemize}
    \item {\em The oracle ${\cal O}_{f_v}$ operation of changing the phase of the state representing information about $v_k$, for which $v_k=v$ is performed.}
    For a binary sequence $v\in\{0,1\}^l$  define the Boolean function $f_v : \{0,1\}^l \rightarrow \{0, 1\}$ by the condition: $f_v(x)= 1$ if and only if $x=v$.

    Let $\ket{x}$ denotes the basic state corresponding to the element $x$, which is one of $v_k$. In this case, $\ket{x}$ is the $l$ - qubit basis state. Oracle ${\cal O}_{f_v}$ performs the following three actions:
  
    \begin{enumerate}
        \item Application of the $H$ operator on the auxiliary ($\log{n} + l+1$)-th qubit $\ket{1}$:
        \[
    \ket{x}\otimes\ket{1}  \quad \xrightarrow{I^{\otimes{\log{n}+l}}\otimes H} \quad
     \ket{x}\otimes \frac{1}{\sqrt{2}}(\ket{0}-\ket{1})
      \]
      \item Application of the operator $ U_{f_v}$ on the last $l+1$ qubits:
       \begin{align*}
   \ket{x}\otimes \frac{1}{\sqrt{2}}(\ket{0}-\ket{1}) & \quad \xrightarrow{I^{\otimes{\log{n}}} \otimes U_{f_v}}  \quad 
 \frac{1}{\sqrt{2}}\ket{x}\otimes(\ket{0\oplus f_v(x)}-\ket{1\oplus f_v(x)}) = \\
 &= (-1)^{f_v(x)}\ket{x}\otimes
 \frac{1}{\sqrt{2}}(\ket{0}-\ket{1}) \\
      \end{align*}
       \item Application of the $H$ operator on the auxiliary ($\log{n} + l+1$)-th qubit $\ket{1}$:
        \begin{align*}
         (-1)^{f_v(x)}\ket{x}\otimes
 \frac{1}{\sqrt{2}}(\ket{0}-\ket{1})
 & \quad \xrightarrow{I^{\otimes{\log{n}+l}}\otimes H} \quad (-1)^{f_v(x)}\ket{x}\otimes\ket{1}
        \end{align*}
   \end{enumerate}

Note that the auxiliary qubit at the end restores its value (by repeated application of the Hadamard transformation) to the $\ket{1}$ state.
    
\item {\em  Inversion operation. }
${\cal D}$ = 2${\cal R}$ - $I^{\otimes \log{n}}$ -- operator applied on the first $n$ qubits. The $\cal{R}$ operator is given in matrix form:
    \[{\cal R}=\frac{1}{\log{n}}\left(\begin{array}{cccc}1 & 1 & \dots & 1\\1 & 1 & \dots  &  1\\ \vdots & \dots & \ddots & \vdots\\ 1
& 1& \dots & 1 \end{array}\right)\]

\end{itemize}
 {\em Getting the result of computation (the output of the ${\cal A}1$).}  After running the macro steps above $\frac{\pi}{4}\sqrt{n}$ times, we measure the first $\log{n}$ qubits in the computational basis. The resulting $\log n$ bits are interpreted as a binary representation of the required index $k$.

\subsection{ Algorithm ${\cal A}$. }

The algorithm consists of two sequentially working parts:
\begin{itemize}
    \item First part: preparing the initial state based on the dictionary $V(string,m)$.
    \item Second part: reading the search word $w$ and searching for its occurrence in the vocabulary.
\end{itemize}

We emphasize that the algorithm ${\cal A}$ has two different input sets: $V(string,m)$ and $w$. These sets are fed to the first and second parts of the algorithm ${\cal A}$, respectively.

  {\em Notations. }
\begin{itemize}
\item     For a binary string $w$ of length $m$, denote by $a(w)$ the number represented by $w$, $0\le a(w)\le 2^m-1$.

Given a number $a\in\{0, \dots, 2^m-1\}$, denote by $bin(a)$ its binary representation of length $m$. 

\item For the vocabulary $V(string,m)$, let $V_p$ denote the following vocabulary
\[
V_p = \{v(w_0), \dots, v(w_{n-1})\},
\]
where $v(w_k)=bin(r(w_k)_p)$ and $r(w_k)_p$ is the $p$-remainder of $a(w_k)$. That is, $a(w_k)= cp+r(w_k)_p$, where $c\ge 0$ and $r(w_k)_p\in \{0,\dots , p-1\}$.
    \item Denote by $P_d = \{p_1, \dots, p_d\}$ the set of first $d$ primes.
\end{itemize}
 
\subsubsection*{Description of the algorithm ${\cal{A}}$.} 

\begin{quote}
 \item {\bf Input:} 
 
For the first part: Vocabulary $V(string,m)$. 

For the second part: Binary word $w$ of length $m$.  
\item {\bf Output:} The index $k$, which is interpreted as an index such that $w=w_ k$.
\end{quote}

That is, ${\cal{A}}$ implements the mapping
\[
{\cal{A}}: V(string,m),  w \longmapsto k.
\]

{\bf The first part of the algorithm } is to prepare the initial state from the vocabulary $V(string,m)$:

\begin{itemize}

\item First, the algorithm makes a classical random choice: a prime number $p$ is uniformly and randomly selected from the set $P_d$
and the vocabulary $V_p$ 
\[
V_p = \{v(w_0), \dots, v(w_{n-1})\}
\]
prepared from the vocabulary $V(string,m)$.

\item {Second step  consists of preparation of  quantum state $\ket{string, p}$ composed of $V_p$}
\[
\ket{string, p} =
\frac{1}{\sqrt{n}}\sum\limits_{k=0}^{n-1}\ket{k}
\otimes \ket{v(w_k)}
\otimes \ket{1}. 
\]

\end{itemize}

{\bf The second part of the  algorithm } is to read the input word $w$ and find $k$ such that $w=w_k$:
\begin{itemize}
\item The algorithm reads the input word $w$ and  prepares $v(w)=bin(r(w)_p)$. 
    \item  The quantum stage  of the algorithm ${\cal{A}}$: quantum procedure  ${\cal{A}}1$ is applied with the input  $\ket{string, p}$ and word $v(w)$. ${\cal{A}}1$ implements the mapping
\[
{\cal{A}}1 : (\ket{string, p}, v(w)) \longmapsto k.
\]
to the state $\ket{string, p}$ and the search word $v(w)$.
The number $k$ is the result of measuring the first $\log{n}$ qubits. The number $k$ is declared as the required number of the word $w_k$ such that $w=w_k$.

\end{itemize}

\subsection{Characteristics of the  algorithm $\cal A$}

The following theorem describes the characteristics of the main part of the $\cal A$ algorithm (searching for the $w$).

\begin{Theorem}\label{th_a2} For a vocabulary $V=V(string,m)=\{w_0, \dots, w_{n-1}\}$ of words of length $m$, for a word $w$ of length $m$ algorithm $\cal A$ solves the problem of finding index $k$ such that $w=w_k$ with the following characteristics. 

For an arbitrary integer $c\ge 3$, for an integer $d=cnm$ it is true that 
\begin{align*}
S^{{\cal A}} (n,m) & =  O(\log{n} + \log{m}), \\
Q^{{\cal A}} (n,m)  & =  O(\sqrt{n}), \\
Er^{\mathcal{A}}(V, w) & \leq  \frac{1}{c} + \frac{1}{n}. 
\end{align*}
\end{Theorem}

The proof of Theorem \ref{th_a2} is given in the next section.

\section{Proof of Theorem\ref{th_a2} }

\subsection{Space complexity $S_\epsilon^{{\cal A}} (n,m)$.} 

We start with the technical Lemma we need below.

\begin{Lemma}\label{space_v(w)}
    For arbitrary $p\in P_d$, for $d=cnm$, for vocabulary 
\[
V_p = \{v(w_0), \dots, v(w_{n-1})\},
\]
generated from vocabulary $V$ and for a word $v(w)$ formed from the word $w$ of length $m$ it is  true that  
\[
|v(w)| = O(\log{n} + \log m), \qquad  |v(w_k)| = O(\log{n} + \log m).
\]
\end{Lemma}

   {\em Proof.}  Due to the Theorem condition we select the set $P_d$ of the first   $d$ primes,  where $d=cnm$. Due to Chebyshev's  theorem there exist constants $0<a<A$ such that for all $d=1,2,…,$ the $d$-th prime number $p_d$ satisfies the inequalities
\[ ad\ln d< p_d< Ad \ln d.\]
That is, for arbitrary $p\in P_d$ and $r\le p$ it is true that 
\[
bin(r) \le \log (Ad \ln d) = O(\log n +\log m).
\]
\Endproof

To estimate the space complexity characteristic $S_\epsilon^{{\cal A}} (n,m)$, consider the  quantum state $\ket{string, p}$ formed on the basis of $V_p$
\[
\ket{string, p} =
\frac{1}{\sqrt{n}}\sum\limits_{k=0}^{n-1}\ket{k}
\otimes \ket{v(w_k)}
\otimes \ket{1}. 
\]

First. $\log{n}$ qubits are needed to encode $\ket{k}$ basis states. Second. According to the \ref{space_v(w)} Lemma, $O(\log n +\log m)$ qubits are needed to encode $\ket{v(w_k)}$ basis states. Thus, we have $S_\epsilon^{{\cal A}} (n,m) =O( \log{n}+\log m)$.

\subsection{The query complexity $Q^{\cal A}(n,m)$.}
The query complexity $Q^{\cal A}(n,m)$ of algorithm  ${\cal A}$ completely determined by the query complexity $Q^{{\cal A}1}(n,m)$ of the  auxiliary procedure  ${\cal A}1$ when  ${\cal A}1$ takes as input the quantum state $\ket{string,p}$. Recall that $\ket{string,p}$ represents (quantumly) the  vocabulary  $V_p$ for $p\in P_d$ with $d$ satisfying the condition of the theorem.

\begin{Property}\label{pr1} Let $p\in P_d$ with $d=cnm$. Let $\ket{string,p}$ be the quantum state  generated by $V_p$, where $V_p$ itself formed by the vocabulary $V(string,m)$. Let  a word  $v(w)$ formed from $w$. Then  for the  ${\cal A}1(\ket{string,p}, v(w))$ the following is true 
\begin{align}
Q^{{\cal A}1} (n,m)  & = O(\sqrt{n})
\end{align}
\end{Property}
{\em Proof.}
Let us consider the amplitude amplification procedure when the procedure ${\cal A}1$ is applied to the initial state $\ket{string, p}$. Let's put $\ket{string_0}=\ket{string, p}$. Let us denote the following numbers by $\alpha_0$ and $\beta_0$: $\alpha_0$ is the initial amplitude for the required basic state $\ket{k}\ket{v(w_k)}\ket{1}$ such that $v(w_k)=v(w)$, and $\beta_0$ are the amplitudes of all other basic states of the initial state $\ket{string_0}$. 

$\alpha_0= 1/\sqrt{n}=\beta_0=1/\sqrt{n}$ and $\alpha_0^2 + (n - 1)\beta_0^2 = 1$. 

By virtue of the introduced notation, the state $\ket{string_0}$ can be represented in the following form:

$$\ket{string_0} = \alpha_0\sum_{k:v(w_k)=v(w)}\ket{k}\otimes\ket{v(w_k)}\otimes\ket{1} + \beta_0\sum_{k:v(w_k) \neq v(w)}\ket{k}\otimes\ket{v(w_k)}\otimes\ket{1}.$$

After applying successively $j$ times two macrosteps to the initial state $\ket{string_0}$ 

1) The operation of changing the phase of the state representing information about $w_k$, for which $v(w_k)=v(w)$ is performed and 

2) The inversion operations (called in the literature Grover's iteration), the amplitudes of the $(j+1)$-th state of the $\ket{string_{j+1}}$ will be expressed by formulas (see, for example, \cite{boyer1998tight} for a detailed technical justification for the effects of operators 1), and 2) on $\ket{string_j}$ states):
   
$$\alpha_{j+1} = \frac{n-2}{n}\alpha_j + \frac{2(n-1)}{n}\beta_j \quad \beta_{j+1} = \frac{n-2}{n}\beta_j - \frac{2}{n}\alpha_j.$$

We have:

$$\ket{string_{j+1}}=
\alpha_{j+1}\sum_{k:v(w_k)=v(w)}\ket{k}\otimes\ket{v(w_k)}\otimes\ket{1} + \beta_{j+1}\sum_{k:v(w_k) \neq v(w)}\ket{k}\otimes\ket{v(w_k)}\otimes\ket{1},$$
where (recall that we are considering the case when $k$ for which $v(w_k)=v(w)$ is the only number).

$$\alpha_{j+1}^2 + (n-1)\beta_{j+1}^2 = 1$$

Therefore, $\alpha$ and $\beta$ can be given as
$$\alpha_j = \sin{((2j + 1)\theta)}, \quad \beta_j = \frac{1}{\sqrt{n - 1}}\cos{((2j + 1)\theta)}$$

Further, $\alpha_r = 1$ if $(2r + 1)\theta = {\pi}/{2}$. Based on these considerations, the optimal number $r$ of iterations of the search algorithm $r = {(\pi - 2\theta)}/{4\theta}$ is determined.

\cite{boyer1998tight} shows that the probability of getting an incorrect result does not exceed ${1}/{n}$ if you run $[{\pi}/{(4\theta)}]$ of Grover's iterations in succession.
If $n$ is a sufficiently large number, then $\theta \approx \sin{\theta} = {1}/{\sqrt{n}}$, then
$$r = \frac{\pi}{4}\sqrt{n}.$$

Since one call to the oracle in the algorithm is testing an entire substring, the final value is $Q^{{\cal A}1} (n,m)= O(\sqrt{n}) $. \Endproof

\subsection{The error $Er^{\cal A}(V, w)$}

The proof of the upper bound for the error probability $Er^{{\cal A}}(V, w)$ is based on the following considerations. For $V$ and the desired word $w$, we split the set $P_d$ of primes into good $P_{good}$
 and bad $P_{bad}$. 
 
 A prime number $p\in P$ will be considered as good for the $V$ and $w$ if $r(w)_p\not=r(w_j)_p$ for all $w_j\in V$ such that $ w\not=w_j$.  That is, the vocabulary $V_p$ with $p\in P_{good}$ (the ``good'' vocabulary $V_p$) represents the vocabulary $V$ ``correctly'', and the vocabulary $V_p$ with $ p\in P_{bad}$ (the  ``bad'' vocabulary $V_p$) represents vocabulary $V$ ``wrong''.

 Then the error  probability $Er^{{\cal A}}(V, w)$ of the algorithm ${\cal A}$ can be estimated from above as follows 
\begin{eqnarray}\label{error_A}
 Er^{\cal A}(V,w)&\le & Pr_{bad} + (1- Pr_{bad})Er^{{\cal A}1}\nonumber \\
 &\le & Pr_{bad} + Er^{{\cal A}1},       
\end{eqnarray}

where $Pr_{bad}$ is the probability of choosing $p\in P_{bad}$ and $Er^{{\cal A}1}$ is the error of ${\cal A} 1$ when ``good'' dictionary $V_p$ is chosen for procedure ${\cal A}1$.
 
 Note that, if a bad $p$ occurs, it is possible to obtain the correct result when applying the ${\cal A}1$ procedure. However, considering all continuations of the procedure ${\cal A}1$ as erroneous for bad $p$, we only increase the error probability. 
 
In the remainder of the proof we estimate the $Pr_{bad}$ and $Er^{{\cal A}1}$ components of the sum (\ref{error_A}).

\begin{Property}\label{pr_bad} For  $d=cnm$ the following is true 
\[
Pr_{bad}\le  \frac{1}{c}.
\]
\end{Property}
{\em Proof.}
Observe that $p$ is selected uniformly at random from $P_d$. So, 
$Pr_{bad}=|P_{bad}|/|P_d|$.  To estimate the $|P_{bad}|$ consider the following. For a pair of distinct numbers $a_1,a_2\in\{0,\dots 2^m-1\}$, we denote by $P_{a_1,a_2}$, the set of primes $p\in P_d$ such that $a_1  \equiv  a_2 \pmod{p}$. For the  vocabulary $V$ and the desired sequence $w$, we have that 
\begin{equation}\label{p_bad_1}
 P_{bad}= \bigcup_{v\in V}P_{a(w),a(v)},  
\end{equation}
Note that for an arbitrary pair of distinct numbers $a_1,a_2\in\{0,\dots , 2^m-1\}$ the following is true
\begin{equation}\label{p_bad_2}
 |P_{a_1,a_2}|\le m.   
\end{equation}

The idea of proving the inequality (\ref{p_bad_2}) is as follows. The number $a=|a_1-a_2|$ does not exceed $2^m$. Therefore, less than $m$ different primes can divide $a$. For details of the proof (\ref{p_bad_2}), see Lemma 7.4 $\cite{RA_1995}$.

Finally combining (\ref{p_bad_1}) and (\ref{p_bad_2}) we have that 
\[
Pr_{bad} \le \frac{nm}{cnm}= \frac{1}{c}.
\]
\Endproof

Now let's estimate the second term of the sum (\ref{error_A}). We estimate the $Er^{{\cal A}1}$ in the condition when we apply the procedure ${\cal A}1$ for the state $\ket{string,p}$.  The state $\ket{string,p}$ represents (quantumly) the  vocabulary 
\[
V_p = \{v(w_0), \dots, v(w_{n-1})\},
\]
for $p\in P_{good}$. That is, we are in the case  where $V_p$ represents the vocabulary $V$ ``correctly''.  This situation satisfies the condition of Grover's search algorithm.  So, immediately we have that

\begin{align}\label{error_A1}
Er^{{\cal A}1} &\le \frac{1}{n}. 
\end{align}

Now, finally,   the probability $Er^{{\cal A}}(string, w)$ is estimated based on (\ref{error_A}) and Property \ref{pr_bad} and (\ref{error_A1}) as follows:

\begin{align*}
Er^{{\cal A}}(string, w) &  \leq \frac{nm}{cnm} +  \frac{1}{n}
   = \frac{1}{c} + \frac{1}{n}.    
\end{align*}

Thus, we have that there are few ``bad variants'' of processing the pair $string$ and $w$ by the ${\cal A}$ algorithm --- their share does not exceed $1/c$ of the total number of possible processings. In the ``good case'' of processing the pair $string$ and $w$, the probability of erroneous processing is bounded from above by the error probability of the procedure ${\cal A}1$, which is bounded by $1/n$.

\section{Generalization}
\label{u_h}

In this section, we present the ${\cal A}2$ algorithm, which is a generalization of the ${\cal A}$ algorithm in terms of universal family of hash functions. 

The concept of universal hashing is defined in \cite{carter1979universal} and has been discussed in sufficient detail in a number of papers, see, for example, \cite{stinson1991universal} and \cite{stinson1994combinatorial}.  The family ${\cal F}=\{f_1, \dots , f_d\}$ of  ($m,l$)-functions $f : \{0,1\}^m\to \{0,1\}^l$ for $l<m $ is called a universal family of hash functions  if for some $\epsilon\in [0,1]$ and for an arbitrary pair $v,w \in \{0,1\}^m$

\[
\frac{|F_{v,w}|}{|{\cal F}|}\le \epsilon, 
\]
where $F_{w,v}=\{f\in {\cal F}: f(v)=f(w)\}$.

For our purposes, we extend the definition of the universal family of hash functions as follows.
\begin{Definition}

 A family ${\cal F}=\{f_1, \dots , f_d\}$ of ($m,l$)-functions 
\[ f: \{0,1\}^m\to \{0,1 \}^l\] 
 for $n\ge 1$ and $\epsilon\in [0,1]$ will be called a {\em strongly $(n,\epsilon)$-universal family of hash ($m,l$)-functions} if for each $n$-subset $Set=\{v_1,\dots , v_n \}$ of the set $\{0,1\}^m$ and an arbitrary word $w\in \{0,1\}^m$
\[
\frac{|F_{Set,w}|}{|{\cal F}|}\le \epsilon, 
\]
where $F_{Set,w}= \bigcup_{v\in Set} F_{v,w}$.

\end{Definition}

Note that a strongly $(1,\epsilon)$-universal family of hash functions is an $\epsilon$-universal family of hash functions in the standard sense.

\paragraph{Example of a Strongly $(n,\epsilon)$-universal family of hash functions. }

An example of a strongly $(n,\epsilon)$-universal family of ($m,l$)-functions is the set 
\[{\cal F}=\{f_1, \dots , f_d\} \] 
of following  functions. For $j\in\{1,\dots, d\}$
($m,l$)-function 
$f_j : \{0,1\}^m \to \{0,1 \}^l$ is determined by the $j$-th prime number $p_j$ as follows: 
\[f_j(w)=bin(r(w)_{p_j}). \]
Here $r(w)_{p_j}$ is the  remainder  $r$ when $a(w)$ is divided by a prime $p_j$ and  $bin(r(w)_{p_j})$ is the binary presentation of the number  $r(w)_{p_j}$.  Clearly we have that $l\le \log p_d$.  

The family $\cal F$ satisfies the following property. 

\begin{Property}\label{freivalds_hash}

   For  $c\ge 3$ and $d=cnm$ the set ${\cal F}=\{f_1, \dots , f_d\}$  of   ($m,O(\log (nm))$)-functions forms a family that  is strongly $(n ,1/c)$-universal family of hash ($m,O(\log (nm))$)-functions. 
   
\end{Property}
{\em Proof.} For the proof see the  Property \ref{pr_bad} above. \Endproof

We now present the ${\cal A}2$ algorithm -- generalization of the ${\cal A}$ algorithm -- in terms of a strongly $(n,\epsilon)$-universal family of hash functions.

{\bf Algorithm ${\cal A}2$.}

The algorithm consists of two sequentially working parts:
\begin{itemize}
    \item First part: preparing the initial state based on the dictionary $V(string,m)$.
    \item Second part: reading the search word $w$ and searching for its occurrence in the dictionary.
\end{itemize}

We emphasize that the algorithm ${\cal A}$ has two different input sets: $V(string,m)$ and $w$. These sets are fed to the first and second parts of the algorithm ${\cal A}$, respectively.

\begin{quote}
\item {\bf Input:} 

For the first part: Vocabulary $V(string, m)=\{w_0, \dots, w_{n-1}\}$ composed of binary words of length $m$ from the $string$. 

For the second part: Binary word $w$ of length $m$.  
\item {\bf Output:} The index $k$, which is interpreted as an index such that $w=w_ k$. 
\end{quote}

{\bf The first part of the algorithm (preparing the initial state):}
\begin{itemize}
\item The first stage of the algorithm -- classical: 
The function $f$ is chosen equiprobably from the set ${\cal F}$.

\item {Second stage (preparation of the quantum state)}:
For the function $f\in{\cal F}$ and vocabulary $V(string,m)$ the algorithm forms vocabulary 
\[
V_f =\{ f(w_0) \dots, f(w_{n-1})\}.
\]

Then, based on $V_f$,  the following quantum state is generated:

\[
\ket{string, f} =
\frac{1}{\sqrt{n}}\sum\limits_{k=0}^{n-1}\ket{k}
\otimes \ket{f(w_k)}
\otimes \ket{1}.
\]
\end{itemize}

{\bf The second part of the algorithm (searching for the $w$):}

\begin{itemize}

\item  The third stage (quantum) of the algorithm ${\cal{A}}2$: 

quantum procedure  ${\cal{A}}1$ is applied with the input $\ket{string, f}$ and word $f(w)$. ${\cal{A}}1$ implements the mapping
\[
{\cal{A}}1 : (\ket{string, f}, f(w)) \longmapsto k.
\]
to the state $\ket{string, f}$ and the search word $f(w)$.

The number $k$ is the result of measuring the first $\log{n}$ qubits. The number $k$ is declared as the required number of the word $w_k$ such that $w=w_k$.

\end{itemize}

Now we have the following statement -- a generalization of the Theorem \ref{th_a2} for the ${\cal A} 2$ algorithm based on ${\cal F}=\{f_1, \dots , f_d\}$ strongly $(n, \epsilon)$-universal family of ($m,l$)-functions.

\begin{Theorem}\label{th_a2+} For a vocabulary  $V=V(string,m)=\{w_0,\dots,w_{n-1}\}$ of $n$ words of length $m$, for a word $w$ of length $m$ algorithm $ {\cal A} 2$  solves the problem of finding an index $k$ such that $ w=w_k$, with the following characteristics

\begin{align*}
S^{{\cal A}2} (n,m) & =  O(\log{n} + l), \\
Q^{{\cal A}2} (n,m)  & =  O(\sqrt{n}), \\
Er^{{\mathcal A}2}(string, w)) & \leq  \epsilon + \frac{1}{n}. 
\end{align*}
\end{Theorem}

{\em Proof. } The proof of Theorem \ref{th_a2+} repeats word for word the proof of Theorem \ref{th_a2} with only one amendment: it is based on the general family of ${\cal F}$ strongly $(n, \epsilon)$-universal  ($m,l$)-functions instead of hash functions functions of a specific family from the Property \ref{freivalds_hash}. 
\Endproof 

\paragraph{Comment.} Note that such generalization  (algorithm ${\cal A}2$)  works effectively  when the length $l$ of hashes is small. As the algorithm  $\cal A$ shows, such algorithms exist -- for the algorithm  $\cal A$ the parameter $l=O(\log (nm))$, see Property \ref{freivalds_hash}.  This provides an upper bound
\[
S^{\cal A} = O(\log n +\log m). 
\]

\section{Conclusion. }

The paper presents a hybrid classical-quantum algorithm ${\cal A}$ and its generalization - the quantum algorithm ${\cal A}2$ for finding the occurrence of the word $w$ in the text $string$. The problem naturally reduces to searching for a word in the vocabulary  $V(string)$ formed from $string$.

The quantum part of algorithms ${\cal A}$ and ${\cal A}2$ presented by the auxiliary quantum procedure ${\cal A}1$. The quantum procedure ${\cal A}1$ is essentially Grover's quantum search algorithm. Here we use the original conditions for applying Grover's algorithm, namely, we consider that in the text $string$ there is (necessarily) a unique occurrence of the word $w$. It must be said that the calculation of query complexity depends on the direct implementation of the algorithm. In this work, we consider the oracle as an appeal to the whole substring, not to its individual bits.

The main declared result in saving the number of qubits used is achieved in the ${\cal A}$ and ${\cal A}2$ algorithms.
The ${\cal A}$ and ${\cal A}2$ algorithms use hash family techniques to save quantum space. The ${\cal A}$ algorithm is based on a specific family of hash functions. This specific family of hash functions is known as Freivald fingerprints (see, for example, the book \cite{RA_1995}, chapter 7 for more information).
The ${\cal A}2$ algorithm is the generalization of the  ${\cal A}$ for an arbitrary strongly $(n,\epsilon)$-universal family of hash functions.

It is important to note that in this paper, as well as in the cited paper \cite{ramesh2003string}, the algorithms are applied to the quantum state (initial state), which is prepared in advance based on the analyzed text $string$ (on the Vocabulary $V(string)$ that is formed from the $string$). The preparation of the initial state requires preliminary work, but this work is not taken into account in the algorithms. Note that the problem of preparing the initial state, as a special problem, is beginning to be discussed in the community dealing with quantum information search. In particular, the authors of \cite{macaluso2020quantum} drew attention to the problem of the complexity of preparing the initial state.

Finally, once again, we note that in this paper we considered the case when it is known in advance that the desired substring occurs in the text exactly once. The problem can be expanded if the number of occurrences is greater than 1 and is known in advance, and also if the number of occurrences of the substring is not known in advance. An estimate of the time and space complexity of Grover's search algorithm in these cases is described in \cite{boyer1998tight}.

\bibliographystyle{unsrt}

\end{document}